\documentclass[prl,twocolumn,superscriptaddress]{revtex4}
\usepackage{amsmath,amssymb,graphicx,color,bbold}
\DeclareMathOperator\Tr{Tr}
\def\ket#1{{\lvert#1\rangle}}
\def\XX{\textit{XX}}
\def\XXZ{\textit{XXZ}}

\begin{document}
\title{Purification and many-body localization in cold atomic gases}

\author{Felix Andraschko}
\affiliation{Department of Physics and Research Center OPTIMAS,
  Technical University Kaiserslautern, D-67663 Kaiserslautern,
  Germany}
\affiliation{Department of Physics and
  Astronomy, University of Manitoba, Winnipeg, Canada R3T 2N2}
\author{Tilman Enss} 
\affiliation{Institut f\"ur
  Theoretische Physik, Universit\"at Heidelberg, D-69120 Heidelberg,
  Germany}
\author{Jesko Sirker}
\affiliation{Department of Physics and Research Center OPTIMAS,
  Technical University Kaiserslautern, D-67663 Kaiserslautern,
  Germany}
\affiliation{Department of Physics and
  Astronomy, University of Manitoba, Winnipeg, Canada R3T 2N2}
\begin{abstract}
  We propose to observe many-body localization in cold atomic gases by
  realizing a Bose-Hubbard chain with binary disorder and studying its
  non-equilibrium dynamics. In particular, we show that measuring the
  difference in occupation between even and odd sites, starting from a
  prepared density-wave state, provides clear signatures of
  localization. As hallmarks of the many-body localized phase we
  confirm, furthermore, a logarithmic increase of the entanglement
  entropy in time and Poissonian level statistics. Our numerical
  density-matrix renormalization group calculations for infinite
  system size are based on a purification approach which allows to
  perform the disorder average exactly, thus producing data without
  any statistical noise, and with maximal simulation times of up to a
  factor $10$ longer than in the clean case.
\end{abstract}
\maketitle

Cold atomic gases with or without optical lattices are an ideal
platform to realize model Hamiltonians of strongly correlated quantum
systems by offering an unprecedented control over the microscopic
parameters \cite{BlochDalibard}. Among the many achievements are the
observation of the superfluid to Mott insulator transition for a Bose
gas held in a three-dimensional optical lattice \cite{GreinerMandel},
the realization of the Tonks-Girardeau regime in a quasi
one-dimensional (1D) Bose gas \cite{par04}, and the simulation of the
non-equilibrium dynamics in an almost integrable one-dimensional
quantum system \cite{KinoshitaWenger}.

While these experiments have all been performed on very clean systems,
there is also a tremendeous interest in building quantum simulators
for models with disorder. This interest is sparked by the unavoidable
presence of disorder and impurities in real materials which can lead
to completely new physics, such as the Kondo effect and Anderson
localization \cite{anderson1958}. For cold atomic gases there have
been several different approaches to realize disorder. The first
experiments have employed quasi-periodic lattices or laser speckles to
study Anderson localization in effectively non-interacting Bose
condensates in one and three dimensions
\cite{RoatiDerricoBillyJosseJendrzejewskiBernardWhitePasienski}.  As
an alternative it has been suggested to use the repulsive interactions
between two different species of atoms---with one being effectively
immobile---to obtain a binary disorder potential for the mobile
species on the scale of the lattice constant
\cite{GavishCastinMorrisonKantianRoscildeCirac,
  HorstmannCiracHorstmannDuerr}.  Experimentally, the localization of
bosonic atoms by fermionic impurities has been demonstrated
\cite{OspelkausOspelkaus_localization}.

Theoretically, it has been suggested by Anderson \cite{anderson1958}
that a localized phase might be stable against small interactions, a
result which has been supported by a recent study \cite{basko2006}
leading to a renewed interest in many-body localization (MBL). In
interacting spin chains with a random magnetic field
{drawn from a box distribution}, in particular, a
transition is found between a delocalized, ergodic phase at weak
disorder and a many-body localized (MBL) phase at strong disorder
\cite{oganesyan2007, pal2010, monthus2010canovi2011}.
In an MBL phase, where all many-body eigenstates are localized, a
simple picture emerges: In this case one can separate a chain
\cite{1D_case} into segments of length $\ell\gg\xi$ where $\xi$ is the
localization length. The many-body eigenstates of the whole chain are
then, to a good approximation, product states of the eigenstates in
each segment.  Importantly, projectors onto the eigenstates of a
segment are {\it local} conserved charges with finite support and have
to be included in a generalized Gibbs ensemble \cite{RigolDunjkoPRL}.
This constrains the dynamics and prevents thermalization in any
subsystem \cite{vosk2013, serbyn2013localhuse2013}.  Residual
interactions between segments make these charges quasi-local, i.e.,
contributions to the conserved charge with spatial support on any
length scale $r$ exist, {but are
  suppressed as $\exp(-r)$. The conserved charges thus remain}
relevant for transport and non-equilibrium dynamics
\cite{ProsenSirkerKonstantinidis}. The residual long-range
interactions are also responsible for the observed growth of the
entanglement entropy, $S\sim\ln t$, with time $t$ if the system is
prepared in a product state and evolves in time
\cite{dechiara2006znidaric2008vosk2013dynamicalserbyn2014,
  bardarson2012, vosk2013}. This is one of the hallmarks of an MBL
phase, in contrast to the linear growth in systems without disorder
\cite{bravyi2006} and the extremely slow increase, $S\sim \ln\ln t$,
found for a noninteracting model with bond disorder \cite{igloi2012}.
{Experimentally, however, this new state of matter has
  not yet been detected and the question which kind of system and
  which {\it local observables} are appropriate for this purpose is
  considered as one of the main open problems in this field
  \cite{AltmanVoskReview}.}

In this letter we discuss the possible realization and observation of
many-body localization in a system of cold atoms. Consider two
interacting species of bosons in an optical lattice, with one of them
frozen to form a binary disorder potential for the other, mobile
species \cite{GavishCastinMorrisonKantianRoscildeCirac,
  HorstmannCiracHorstmannDuerr}.  The effective Hamiltonian for the
mobile bosons is
\begin{eqnarray}
\label{Hbos}
H&=&-\frac{J}{2}\sum_j \left(a^\dagger_j a_{j+1} +h.c.\right) \\
&+& \sum_j\left[U n_j(n_j-1)/2 +Vn_jn_{j+1}+JD_jn_j\right], \nonumber
\end{eqnarray}
where $J$ is the hopping amplitude, $n_j=a_j^\dagger a_j$ the local
density, $U$ the onsite, and $V$ the nearest-neighbor interaction. The
effective binary disorder potential $D_j$ is drawn randomly according
to $D_j=\pm D$, and we have {neglected} the trapping
potential.  We consider in the following the case where the mobile
species is prepared in the initial density-wave state
$\lvert\Psi_0\rangle =\lvert 010101\cdots\rangle$. During the ensuing
time evolution under the Hamiltonian \eqref{Hbos} we propose to
measure the difference in occupation between the even and odd sites,
$\Delta n=N^{-1}\sum_j (-1)^j \left(\langle n_j\rangle-1/2\right)$,
where $N$ is the number of lattice sites. Exactly this setup has
already been realized in the clean case, i.e., for a single species of
bosons \cite{TrotzkyChen}. By freezing the immobile bosons into a
quantum state which is close to an equal superposition of Fock states
$|n_1, n_2,\cdots\rangle$ with $n_j\in\{0,1\}$, the time evolved state
is automatically averaged over all binary disorder configurations
\cite{paredes2005, HorstmannCiracHorstmannDuerr}.  This purification
method can also be used for numerical computations and is explained in
detail below.  Our proposal thus combines realizing disorder in
optical lattices using two species of atoms
\cite{OspelkausOspelkaus_localization} with techniques to prepare
initial states and to measure their non-equilibrium dynamics
\cite{TrotzkyChen}. Experimentally, this can be realized, for
instance, with two hyperfine states of $^{87}$Rb atoms loaded into a
state-dependent optical lattice: the wavelength controls the relative
hopping amplitude of both states, while the intraspecies interaction
is tuned by a Feshbach resonance, and the interspecies interaction
depends on the intensity of the laser beam.  In this way, both the
ratio of onsite interaction $U$ to hopping $J$ and the coupling
between mobile and disorder atoms can be tuned independently, and
bimodal disorder with very short-range correlations is realized
\cite{HorstmannCiracHorstmannDuerr}.

Let us first discuss the cases $V=0$ with $U=0$ or $U=\infty$ with
strong disorder $D\gg1$. In this case, each disorder configuration
splits the chain into segments of equal potential $D_i$, which
communicate very little with their neighboring segments due to the
mismatch of the local potential energy and the kinetic energy, $JD\gg
J$: for a given segment, $\Delta n$ has contributions from neighboring
segments of the order of $1/D^2$.  Thus, the segments become isolated
in the limit $D\to\infty$.  The time evolution of the whole chain is
then given by summing up the independent time evolution of open
segments $\Delta n^{(\ell)}(t)$ of varying length $\ell$, weighted by
their probability of occurrence, $p_\ell = \ell/2^{\ell+1}$ with
$\sum_{\ell} p_\ell = 1$, {leading to $\Delta
  n^{D=\infty}(t) = \sum_{\ell=1}^\infty p_\ell \Delta n^{(\ell)}(t)$
  both for $U=0$ and $U=\infty$} 
\cite{SupplMat}.  It is easy to see that only segments of odd chain
length contribute to the long-time average with $\overline{\Delta
  n}^{(\ell)}_{\text{odd}} = 1/(2\ell)$, so that $\overline{\Delta n}
= \sum_\ell p_\ell \overline{\Delta n}^{(\ell)} = \frac12 \sum_{\ell
  \text{ odd}} p_\ell/\ell = \frac16$. 

\begin{figure}[ht!]
  \centering
  \includegraphics[width=0.99\columnwidth]{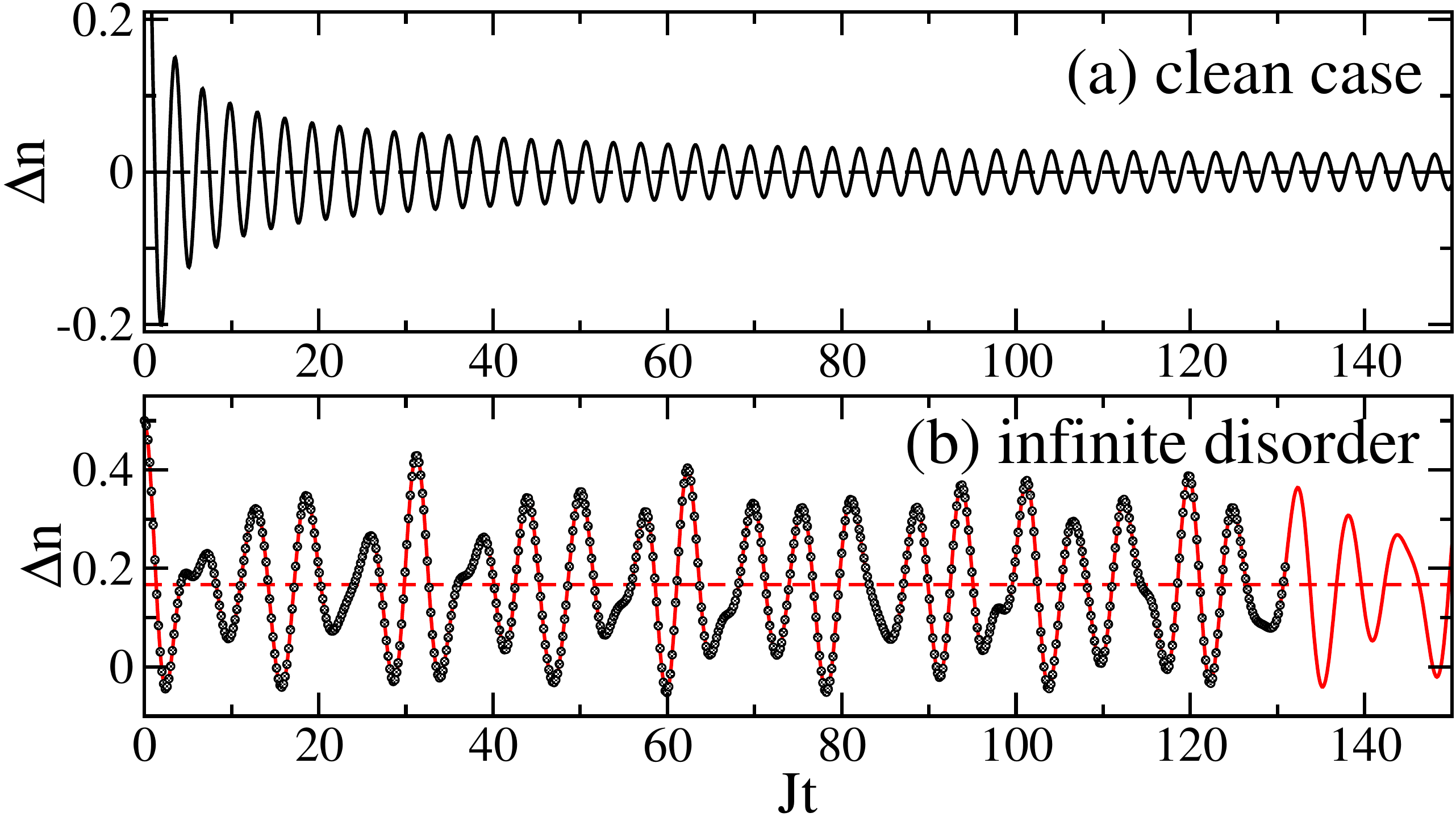}
  \caption{(Color online) $\Delta n(t)$ for {model
      (\ref{Hbos}) with} $V=0$ and $U=0$ or $U=\infty$: (a) $\Delta
    n^{D=0}$ vs (b) $\Delta n^{D=\infty}$ (solid red) with average
    $\overline{\Delta n}=1/6$ (dashed red); symbols are LCRG results.}
  \label{fig:XXdis}
\end{figure}

Fig.~\ref{fig:XXdis}(a) shows the time evolution in the clean case,
where for both $U=0$ and $U=\infty$ one finds $\Delta n^{D=0}(t) =
J_0(2Jt)/2$, with the Bessel function $J_0(x)$
\cite{barmettler2009barmettler2010, enss2012lcrg}.  In comparison, the
time evolution in the strongly disordered case,
Fig.~\ref{fig:XXdis}(b), appears very complicated. Yet, including only
segments up to a maximum length $L$ already gives an approximation
with an exponentially small error $\sim L/2^{L}$ at all times, because
$p_\ell$ decays exponentially.

The model \eqref{Hbos} is difficult to treat numerically because of
the unrestricted local Hilbert space dimension for finite interaction
strength $U$. We will therefore first concentrate on the limit
$U\to\infty$, where numerical methods are very efficient. We will
firmly establish that an MBL phase exists in this limit before
returning to the case of finite $U$ at the end of this letter. For
$U\to\infty$, the above model maps onto the spin-1/2 \XXZ\ chain
\begin{align}
\label{model}
  H = -J\sum_i \bigl( s_i^x s_{i+1}^x + s_i^y s_{i+1}^y - \Delta s_i^z
  s_{i+1}^z - D_i s_i^z \bigr),
\end{align}
with anisotropy $J\Delta=V$ \cite{GiulianoRossini}. $|\Psi_0\rangle$
is then the N\'eel state and $\Delta n=N^{-1}\sum_j (-1)^j \langle
s^z_j\rangle$ is the staggered magnetization. For $\Delta=0$, the
dynamics is again given by Fig.~\ref{fig:XXdis}(a) in the clean case
and by Fig.~\ref{fig:XXdis}(b) in the strongly disordered case. For
alkali atoms, {$V\ll J$}, so that a realization of the
\XXZ\ model with substantial $\Delta$ is not easily achievable.  We
note, though, that the isotropic Heisenberg chain, $\Delta=1$, has
recently been realized using two boson species, and that the
non-equilibrium dynamics has been studied with single-site
addressability \cite{FukuharaKantian}.  Furthermore, longer-range
interactions are also present if dipolar gases or polar molecules are
used \cite{Hazzard}.

To simulate the non-equilibrium dynamics of the Bose-Hubbard model
\eqref{Hbos} and the \XXZ\ model \eqref{model} for all interaction and
disorder strengths and to perform an exact disorder average we use the
light cone renormalization group (LCRG) \cite{enss2012lcrg}. The LCRG
algorithm is a variant of the density-matrix renormalization group
(DMRG) technique \cite{white1992} based on the Lieb-Robinson bounds
\cite{lieb1972}: a local measurement at time $t$ is affected only by
the degrees of freedom within its light cone.  The opening angle of
the light cone, or spreading velocity, is determined by model
parameters.  For lattice models with short-range interactions, the
time evolution operator $\mathcal U=\exp(-iHt)$ has a Trotter-Suzuki
decomposition with a checkerboard structure: within the LCRG, a light
cone out of the infinite checkerboard is sufficient to compute the
time evolution of local observables in an infinite system
\cite{enss2012lcrg, AndraschkoSirker}.

\begin{figure}[ht!]
  \centering
  \includegraphics[width=0.99\columnwidth]{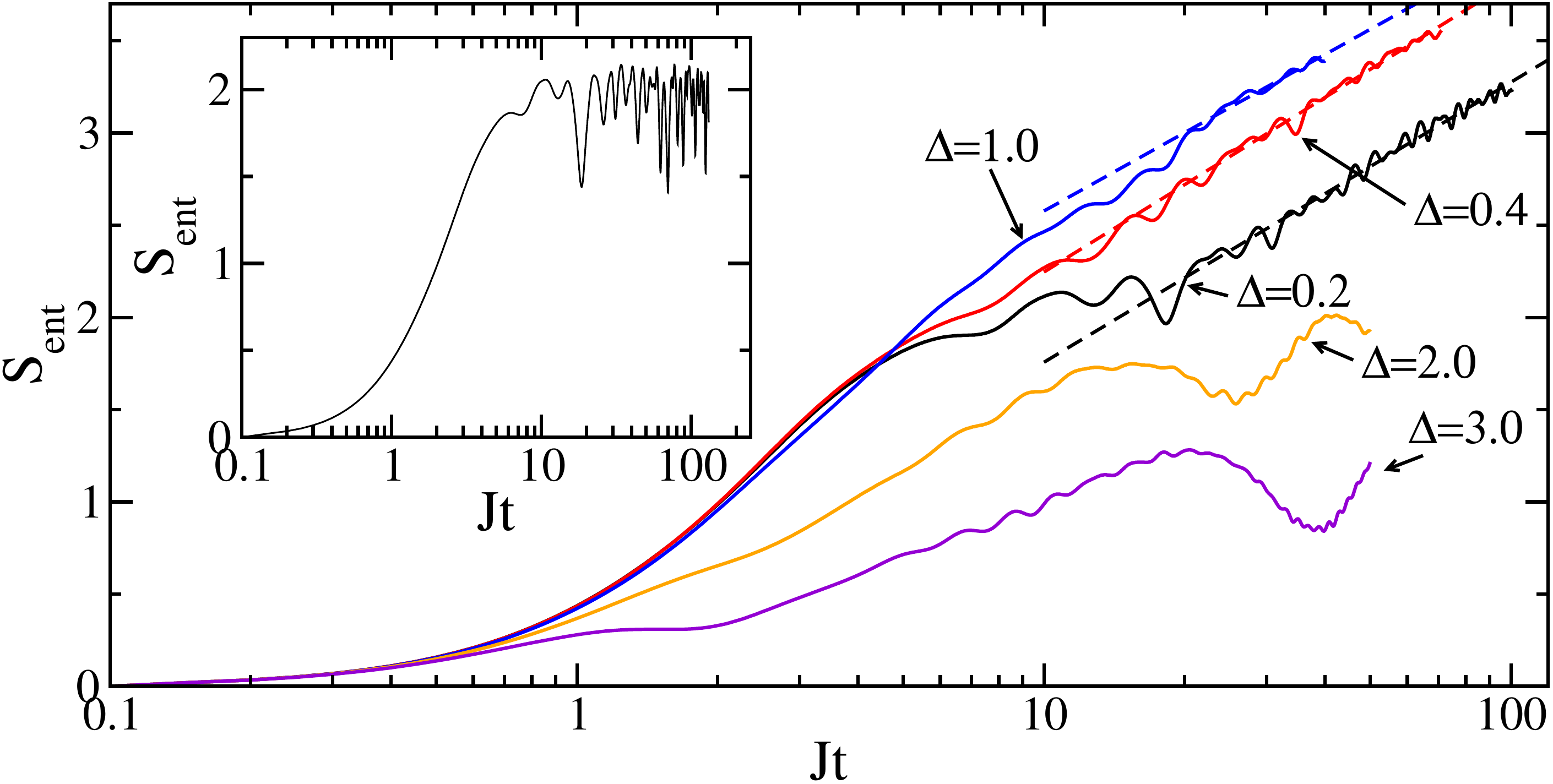}
  \caption{(Color online) $S_\textrm{ent}$ {for the
      \XXZ\ chain \eqref{model}} in the strongly disordered case
    $D=4000$. For small $\Delta$ we find asymptotically
    $S_\textrm{ent}(t)\sim\ln t$ (dashed lines are fits for $t>20$).
    Inset: $S_\textrm{ent}(t)$ saturates for $\Delta=0$ and infinite
    disorder.}
  \label{fig:entanglement}
\end{figure}

To treat disorder, one straightforward possibility {for a {\it finite
    system}} is to compute the time evolution for one particular
disorder configuration, and then repeat the calculation for many
different configurations to obtain the disorder average.  Here, we
instead use purification {for an {\it infinite system}} in order to
perform the full disorder average in a single run \cite{paredes2005},
at the expense of enlarging the Hilbert space.  Specifically, for the
\XXZ\ chain, an ancilla spin-$1/2$, $\vec s_{i,\text{anc}}$, is added
to each lattice site with an Ising coupling, $D_i s_i^z \mapsto
2Ds_i^z s_{i,\text{anc}}^z$. The state of $s_{i,\text{anc}}^z=\pm1/2$
now determines the local Zeeman field $D_i=\pm D$.  There is no
coupling between different ancilla spins, hence they have no dynamics
and represent static disorder. The time evolution of the disorder
average is given by the evolution from a prepared product state
$\ket{\psi_0}\otimes\ket{\text{dis}}$ in the enlarged Hilbert space of
spins and ancillas, where $\ket{\text{dis}}=\bigotimes_j
(\ket{\uparrow}_j+\ket{\downarrow}_j)/\sqrt{2}$ is the fully mixed
state for the ancillas. The disorder averaged expectation value of an
operator $O$ is then obtained by measuring the expectation value of
the operator $O\otimes \mathbb{1}_{\text{anc}}$ in the enlarged
Hilbert space.  Although the local Hilbert space dimension is doubled,
the LCRG algorithm works even more efficiently for strongly disordered
systems than for clean systems, and real times up to $Jt\sim 100$ are
reached in our simulations, where we keep the truncation error in each
renormalization group step smaller than $10^{-8}$ by dynamically
increasing the number of kept states up to $20000$. Responsible for
these long simulation times is the slow logarithmic growth of the
entanglement entropy, $S_\textrm{ent}$, for $\Delta\neq 0$, see
Fig.~\ref{fig:entanglement}.
Here, $S_\textrm{ent}=-\Tr\rho_B\ln\rho_B$, where $\rho_B$ is the
reduced density matrix obtained by cutting the infinite chain,
$A\otimes B$, of spins {\it and} ancillas in half. Since entanglement
in the static ancillas is mediated by the spins, $S_\textrm{ent}$ has
the same functional dependence on time as the disorder averaged
entanglement entropy of a spin-only system \cite{SupplMat}. The
logarithmic increase for $\Delta\neq 0$ is the same behavior as seen
for the \XXZ\ model with the magnetic fields $D_i$ drawn from a box
distribution \cite{bardarson2012}, and is a hallmark of an MBL phase.
On the other hand, $S_\textrm{ent}$ saturates for $\Delta=0$ and
infinite binary disorder, see the inset of
Fig.~\ref{fig:entanglement}. The latter behavior can be easily
understood by noting that $S_\textrm{ent}$ for a block of size
$n\leq\ell$ of a finite chain segment of spins and ancillas with
length $\ell$ is bounded, $S_\textrm{ent}\leq n\,\ln 4$.  Since
$p_\ell$ decreases exponentially, a strict bound for
$S_{\textrm{ent}}$ at {all} times exists
\cite{SupplMat}. This is different from the case of strong bond
disorder, where $S_\text{ent}\sim \ln \ln t$ \cite{igloi2012}.

\begin{figure}[ht!]
  \centering
  \includegraphics[width=0.99\columnwidth]{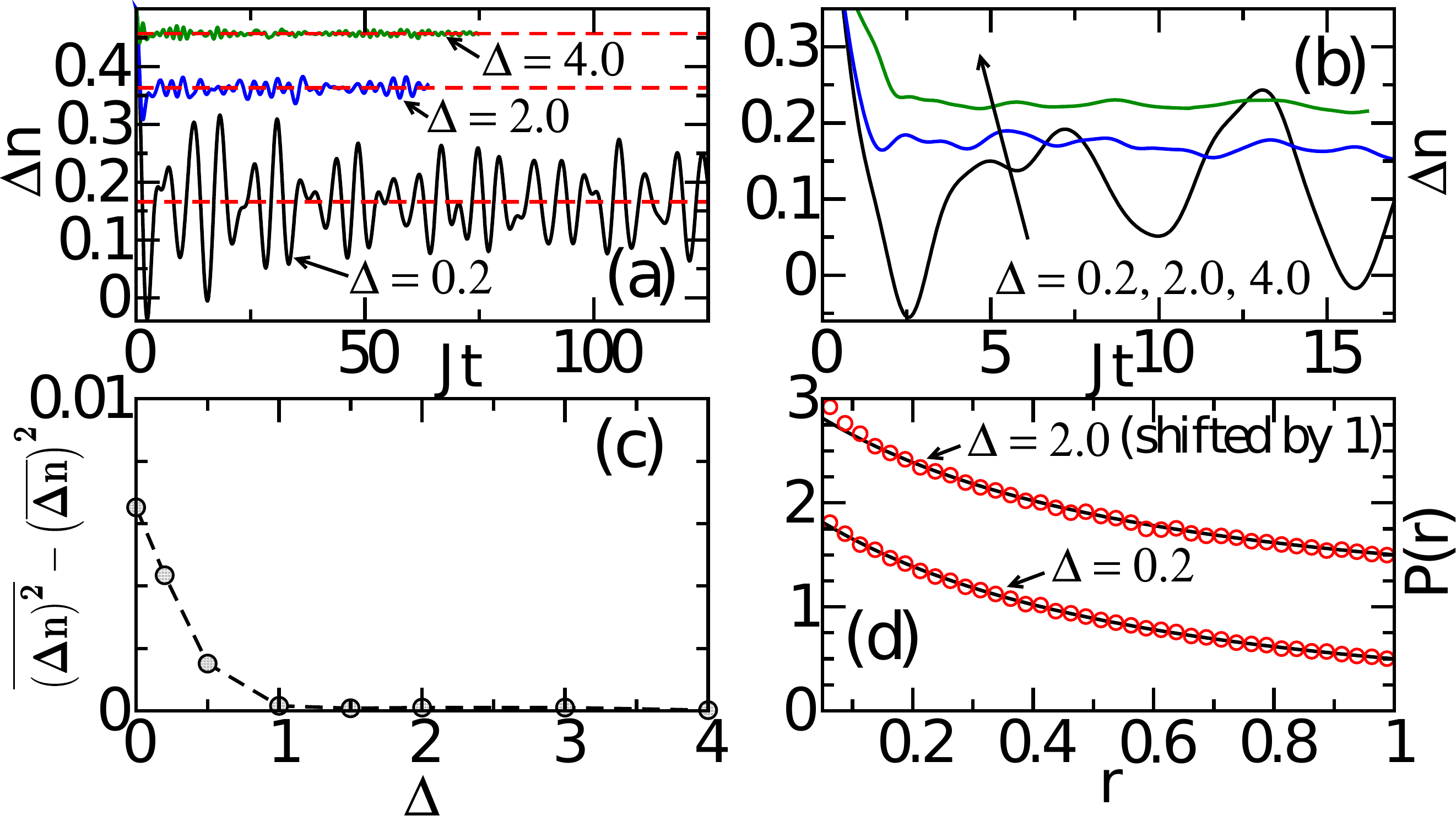}
  \caption{(Color online) {\XXZ\ chain:} (a) $\Delta
    n(t)$ for $D=4000$ with averages (dashed lines). $D=1.5$: (b)
    $\Delta n(t)$, (c) {variance} of $\Delta n(t)$
    for $t>5$, and (d) $P(r)$ for chains of length $N=14$ (symbols) and
    $P(r)=2/(1+r)^2$ (solid lines).}
  \label{fig:ms_and_r}
\end{figure}

In Fig.~\ref{fig:ms_and_r}(a,b) we show $\Delta n(t)$ for strong and
intermediate disorder.
In all cases shown, $\Delta n(t)$ does not decay to zero, indicating
that the system does not thermalize. A clear experimental indication
that localization in an interacting system is observed, is the
{strong reduction of the variance of $\Delta n(t)$ with
  increasing $\Delta$}, see Fig.~\ref{fig:ms_and_r}(c).

To further support our findings of an MBL phase for the \XXZ\ model
with binary disorder, we have also calculated the level statistics for
finite chains of up to $N=14$ sites in the $S^z=0$ sector by exact
diagonalization of all $2^N$ possible disorder realizations. In the
integrable \XXZ\ chain without disorder, a full set of local integrals
of motion exists, which allows to completely classify the eigenvalues
by the corresponding quantum numbers. The spectrum is therefore
uncorrelated and the corresponding level statistics Poissonian,
$\mathcal{P}(s)=\exp(-s)$, in terms of the level spacing $s$. Disorder
breaks integrability, so that the level-spacing distribution, if the
many-body states are extended, will follow a Wigner distribution,
$\mathcal{P}(s)=(\pi s/2)\exp(-\pi s^2/4)$. This can also be
understood as a crossover from integrability to quantum chaos
\cite{YurovskyOlshanii}. However, once localization sets in, the
spectrum will again become uncorrelated, because localization creates
new quasi-local conserved charges, leading to a Poissonian level
statistics. In this case chaos is incomplete and the system keeps a
memory of the initial state \cite{YurovskyOlshanii}. If a critical
$D_c\neq 0$ for localization exists, we therefore expect to go from a
Poissonian ($D=0$, integrable) to a Wigner distribution ($0<D<D_c$,
non-integrable and delocalized), and then again back to a Poissonian
($D>D_c$, localized) \cite{SupplMat}. Here, we concentrate on the
regime far from the clean integrable limit.  In order to avoid the
ambiguous definition of an average gap based on a construction of a
continuous density of states from finite size data, we consider the
ratio $r$ between two consecutive gaps of adjacent energy levels as
defined in Refs.~\cite{oganesyan2007, pal2010}.  If the level
statistics is Poissonian, the distribution function of gap ratios
$0\leq r\leq 1$ is given by $P(r)=2/(1+r)^2$.  As shown in
Fig.~\ref{fig:ms_and_r}(d), this is in good agreement with the
numerical results \cite{SupplMat}.

\begin{figure}[ht!]
  \centering
  \includegraphics[width=0.99\columnwidth]{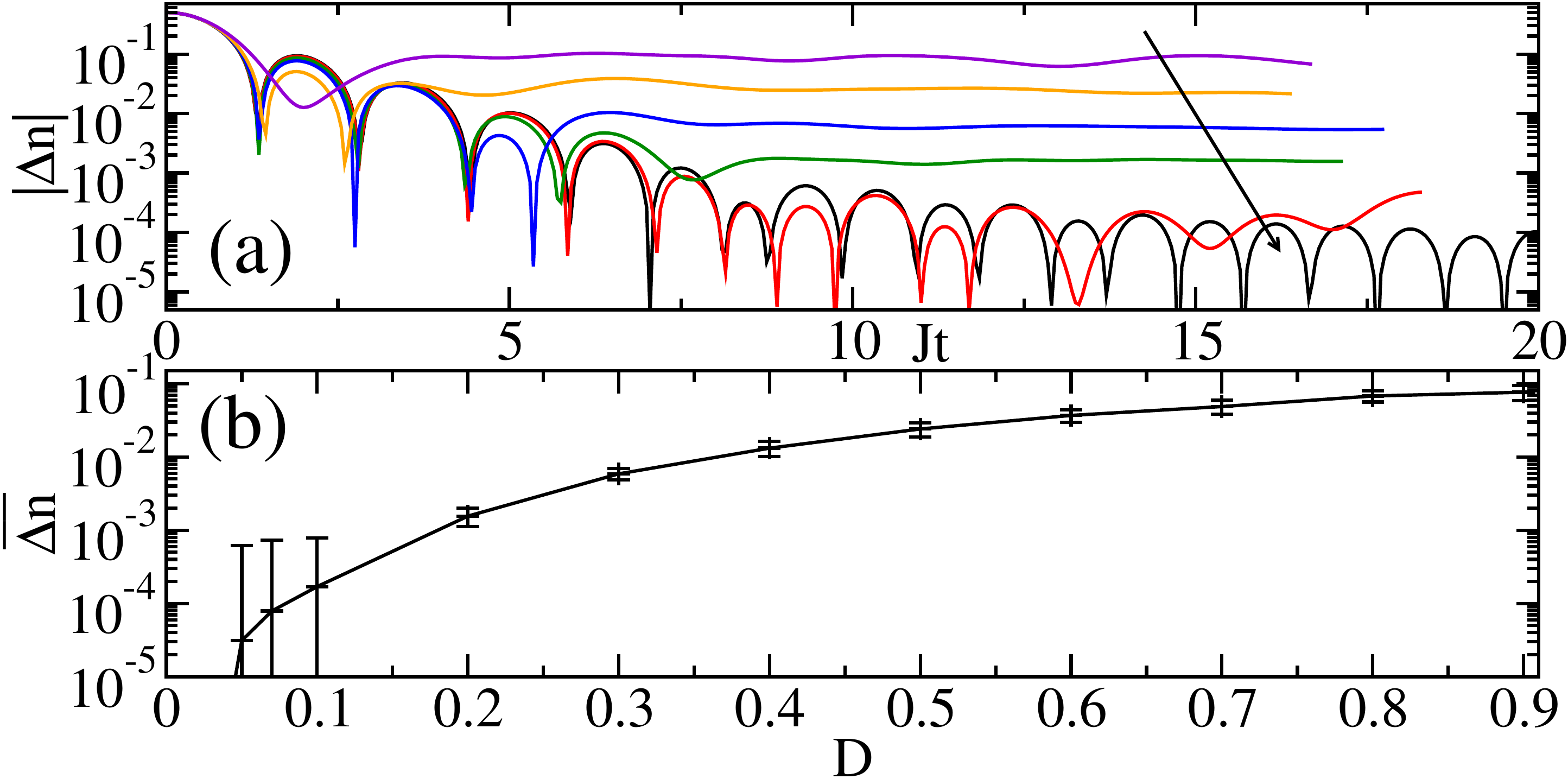}
  \caption{(Color online) (a) $\Delta n(t)$ for the Heisenberg point,
    $\Delta=1$, and disorder strengths $D=0.9,$ $0.5,$ $0.3,$ $0.2,$
    $0.1,$ $0.0$ (in arrow direction). (b) $\overline{\Delta n}$
    obtained by averaging over intervals $[t_\textrm{min},20]$ with
    $t_\textrm{min}>5$ variable (see error bars).}
  \label{fig:HB}
\end{figure}

Of particular interest is the isotropic Heisenberg chain, $\Delta=1$,
which has recently been realized in cold atomic gases
\cite{FukuharaKantian} and approximately describes materials such as
Sr$_2$CuO$_3$ and SrCuO$_2$ \cite{MotoyamaEisaki}. In the latter case,
a doping with non-magnetic impurities, such as Pd, is possible, which
randomly replace the magnetic Cu$^{2+}$ ions \cite{Kojima}. As in the
model \eqref{model}, for strong disorder, the chain then separates into
segments, however, the coupling between the segment ends is now not an
Ising, but a Heisenberg coupling caused by next-nearest neighbor
interactions \cite{SirkerLaflorencieSirkerLaflorencie2}. In such
systems, many-body localization should also occur, preventing local
excitations from spreading. However, phonons complicate the
observation of MBL, leaving cold atomic gases as a particularly clean
and promising realization.  In Fig.~\ref{fig:HB}, results for the
Heisenberg model with binary disorder are shown.
Here, clear signatures for MBL are already seen for small disorder,
with $\Delta n(t)$ quickly approaching a non-zero constant value, see
Fig.~\ref{fig:HB}(a), while $\Delta n(t)$ decays completely in the
clean case \cite{barmettler2009barmettler2010}. By averaging over
different time intervals, we can extract an estimate for the long-time
average $\overline{\Delta n}$ with error bounds, see
Fig.~\ref{fig:HB}(b). The results are consistent with a critical value
$D_c$ for the localization transition which is either zero or finite,
but with $D_c<0.2$. A detailed analysis of the phase diagram,
including the crossover from integrable to (incomplete) chaotic
behavior, will be presented elsewhere
\cite{EnssAndraschkoSirker_prep}.

\begin{figure}[ht!]
  \centering
  \includegraphics[width=0.99\columnwidth]{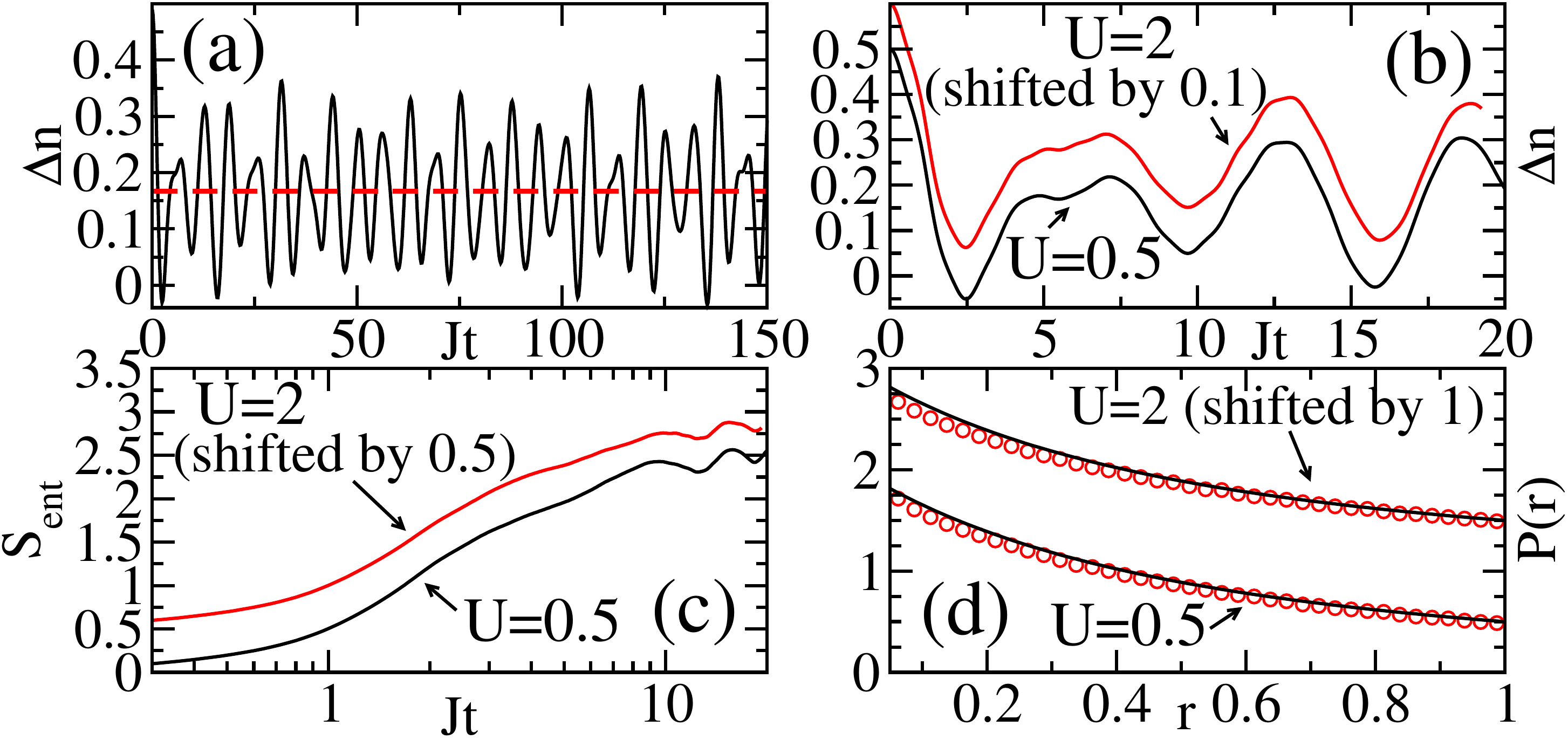}
  \caption{(Color online) Bose-Hubbard model with $V=0$. (a) $\Delta
    n(t)$ for $D=\infty$ and $U=2$ with average (dashed line).
    Disorder strength $D=6$: (b) $\Delta n(t)$, (c)
    $S_\textrm{ent}(t)$, and (d) $P(r)$ for $N=14$ (symbols) and
    $P(r)=2/(1+r)^2$ (solid lines).}
  \label{fig:BH}
\end{figure}

Let us finally return to the full Bose-Hubbard model \eqref{Hbos},
which we have proposed to realize experimentally. We concentrate on
the case of vanishing nearest-neighbor interaction, $V=0$, realized in
alkali atoms. For infinite disorder strength, the system still
separates into decoupled chain segments for any interaction strength
$U$. In this case, we can simulate the dynamics for arbitrary times by
exactly diagonalizing the segments, see Fig.~\ref{fig:BH}(a).  Note
that in the limit $D\to\infty$, $S_\textrm{ent}(t)$ is bounded: there
is no many-body localization in this case. At finite disorder, we
again use the LCRG algorithm to simulate the system. Results for
$\Delta n(t)$ and $S_\textrm{ent}(t)$ are shown in
Fig.~\ref{fig:BH}(b,c), respectively.
Although the simulation time is more limited than in the \XXZ\ case,
$\Delta n(t)$ seems to remain nonzero, while the data for
$S_\textrm{ent}$ are consistent with a logarithmic increase, as
expected in an MBL phase. This is corroborated further by the
distribution function of gap ratios $P(r)$, see Fig.~\ref{fig:BH},
consistent with a Poissonian level statistics.

To conclude, we have proposed to study many-body localization in cold
atomic gases by realizing a Bose-Hubbard model with binary disorder
provided by a second species, and studying its quench dynamics. Both
experimentally as well as in numerical calculations one can make use
of purification to achieve an automatic disorder average.  By
implementing the purification scheme into a DMRG algorithm, we have
shown that the non-equilibrium dynamics can be simulated in a single
run without any stochastic noise, and with simulation times for strong
disorder which are significantly longer than in the clean case, making
DMRG-type algorithms an ideal tool to investigate infinite, disordered
systems.  Both in the Bose-Hubbard model 
as well as in the \XXZ\ chain limit 
we have shown that an MBL phase exists {and can be detected by
  measuring the {\it one-point function} $\Delta n(t)$.}

\acknowledgments F.A. and J.S. acknowledge support by the
Collaborative Research Centre SFB/TR49, the Graduate School of
Excellence MAINZ (DFG, Germany), as well as NSERC (Canada). We are
grateful to the Regional Computing Center at the University of
Kaiserslautern, the AHRP, and Compute Canada for providing
computational resources and support.

\clearpage
\appendix

\onecolumngrid
\setcounter{equation}{0}
\begin{center}
\LARGE{Supplemental Material:\\ ``Purification and many-body
  localization in cold atomic gases''\\ by F.~Andraschko, T.~Enss, and
  J.~Sirker}
\end{center}

We present in Sec.~I additional information about the exact results
for the free models, and in Sec.~II data for the spin-only
entanglement entropy for the \XXZ\ model. In Sec.~III we discuss in
detail the calculation of the level statistics for finite chains
{and in Sec.~IV the LCRG algorithm.}

\section{I. Free models}
\label{XX}
The Bose-Hubbard model with $V=0$ reduces to free bosons for $U=0$ and
to hard-core bosons for $U=\infty$. Similarly, the \XXZ\ chain for
$\Delta=0$ reduces to free fermions.

\subsection{Difference in occupation of even and odd sites}
For the difference in occupation between the even and odd sites,
$\Delta n(t)$, the statistics does not matter in the clean case,
$D=0$, and in all three limits we find
\begin{equation}
\label{clean}
\Delta n^{D=0}(t) = J_0(2Jt)/2 
\end{equation}
where $J_0(t)$ is the Bessel function of the first kind which decays
$\sim 1/\sqrt{t}$ for large $t$. The statistics is also irrelevant for
infinite disorder, $D=\infty$, where the chain splits into decoupled
segments and the full time evolution of $\Delta n^{D=\infty}(t)$ can
be calculated exactly as well. Using a Fourier representation one
obtains
\begin{align}
  \label{eq:mlt}
  \Delta n^{(\ell)}(t) = \frac{1}{2\ell} \sum_{k=1}^\ell \exp\left[
    2iJt \cos\left(\pi \frac{k}{\ell+1}\right) \right]
\end{align}
for a segment of length $\ell$ with open boundary conditions and
constant disorder potential. Here, the contributions with momenta $k$
and $\ell+1-k$ are complex conjugate, so that $\Delta n(t)$ is real.
Initially, $\Delta n^{(\ell)}(t=0)=1/2$ for all segment lengths
$\ell$, and for large segments $\Delta n^{(\ell\to\infty)}(t) \to
\Delta n^{D=0}(t)$ approaches the thermodynamic limit result of the
chain without disorder, Eq.~\eqref{clean}. The time evolution of the
whole chain is then given by summing up the time evolution of segments
of varying length $\ell$, weighted by their probability of occurrence
$p_\ell=\ell/2^{\ell+1}$,
\begin{equation}
   \label{eq:mst}
   \Delta n^{D=\infty}(t) = \sum_{\ell=1}^\infty p_\ell \Delta
   n^{(\ell)}(t)
   = \frac12 \left[ \frac14 \times 1 + \frac28 \times \cos(Jt) 
   + \frac{3}{16} \times (1+2\cos(\sqrt2Jt)) + \dotsm \right],
\end{equation}
where we have displayed $\Delta n^{(\ell)}(t)$ for open chains of
lengths $\ell=1,2,3$ explicitly. The series converges exponentially
fast.

\begin{figure}[ht!]
  \centering
  \includegraphics[width=0.62\columnwidth,clip]{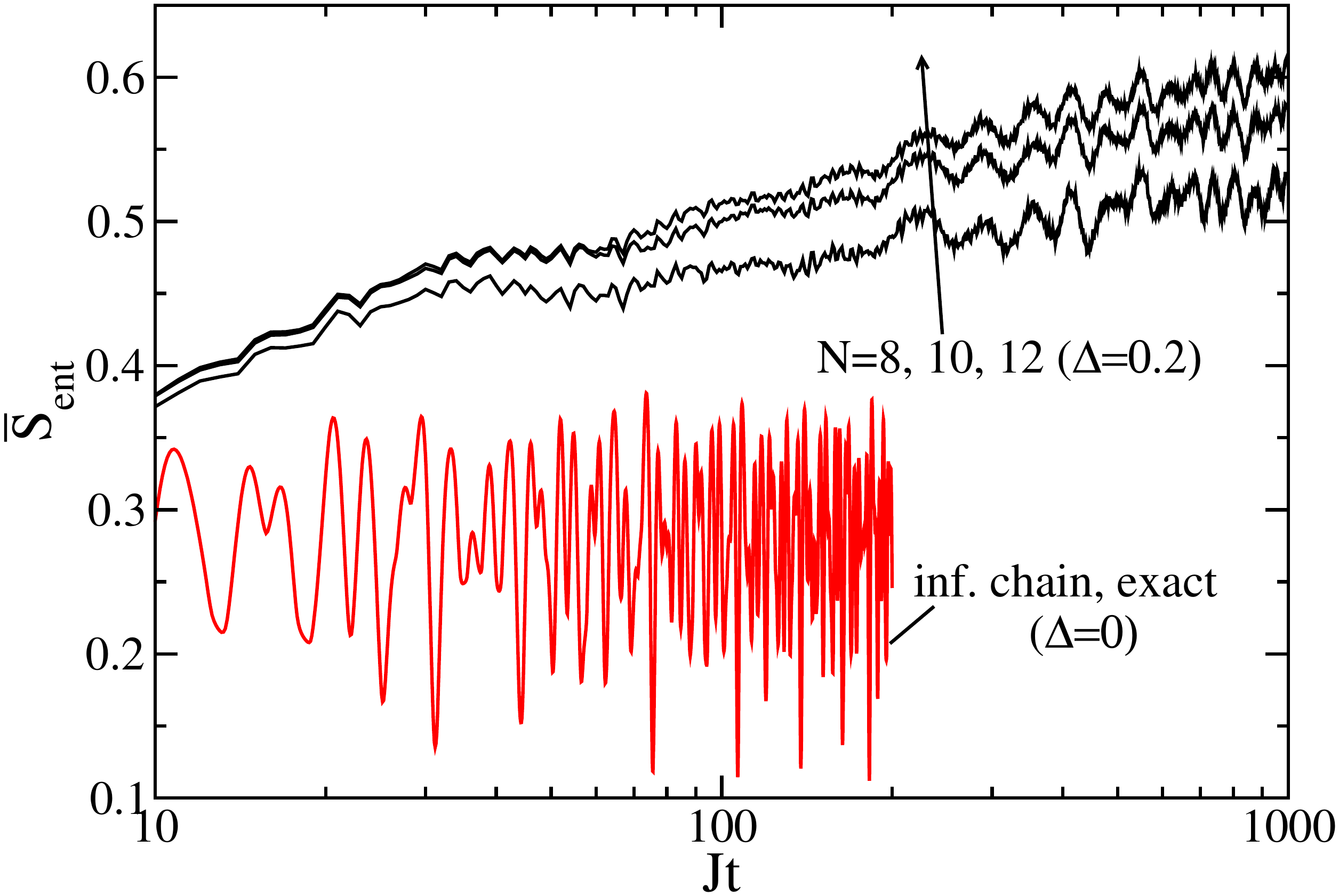}
  \caption{Disorder-averaged entanglement entropy, Eq.~\eqref{Sent1},
    for the \XXZ\ chain with $D=4000$, $\Delta=0.2$, and system sizes
    $N=8,10,12$. For comparison, the exact result for an infinite \XX\
    chain ($\Delta=0$) with infinite disorder is shown---for clarity
    only up to $Jt=200$---where
    $\overline{\mathcal{S}}_{\textrm{ent}}(t)$ is bounded.}
  \label{fig:exactee}
\end{figure}

\subsection{Entanglement entropy}
For the considered models with binary disorder, there are two kinds of
entanglement entropies: one with and one without ancillas. Without
ancillas, the Hilbert space of the system, consisting of spins or
bosons, is written as a direct product,
$\mathcal{H}=\mathcal{H}_A\otimes\mathcal{H}_B$, and the
disorder-averaged entanglement entropy of block $A$ is defined as
\begin{equation}
\label{Sent1}
{\overline{\mathcal{S}}_{\textrm{ent}}(t)=-\frac{1}{2^N}\sum_{\mathcal{C}}\Tr\left\{\rho^\mathcal{C}_A(t)\ln\rho^\mathcal{C}_A(t)\right\}}
\end{equation}
{with
  $\rho^\mathcal{C}_A(t)=\Tr_B\rho^\mathcal{C}(t)$}. Here
$\mathcal{C}$ denotes the $2^N$ configurations of the binary disorder
potential for a chain of length $N$ {and
  $\rho^\mathcal{C}(t)$ the corresponding density matrix.} In the
paper we have used an enlarged Hilbert space consisting of the system
(s) and ancillas (a) with
$\mathcal{H}=\mathcal{H}_{A_s}\otimes\mathcal{H}_{A_a}\otimes
\mathcal{H}_{B_s}\otimes\mathcal{H}_{B_a}$ and the entanglement
entropy shown in Fig.~2 and Fig.~5 of the paper is defined as
\begin{equation}
\label{Sent2}
{S_{\textrm{ent}}(t)=-\Tr\left\{\rho_A(t)\ln\rho_A(t)\right\}}
\end{equation}
{with $\rho_A(t)=\Tr_{B_s}\Tr_{B_a}\rho(t)$.} Starting
from a product state, the entanglement spreads during the time
evolution via the hopping (spin flips) on the system sites, while the
ancillas do not have any dynamics. The asymptotic functional time dependence of both entanglement entropies will therefore be the same.

For the non-interacting cases, the time evolution of the disorder
averaged entanglement entropy
$\overline{\mathcal{S}}_{\textrm{ent}}(t)$ can be calculated
analytically for infinite disorder. The probability that a chain
segment with length $\ell\geq 2$ is cut with $d$ sites left open and
$\ell-d\geq 1$ sites traced over is $1/2^{\ell+1}$.  There are
$\ell-1$ possibilities to cut a segment of length $\ell$ and
\begin{equation}
\label{Sent_an1}
\sum_{\ell=2}^\infty \frac{\ell-1}{2^{\ell+1}} =\frac12 
\end{equation}
while with probability $1/2$ the cut is in between segments in which
case the entanglement entropy is zero. The disorder averaged
time-dependent entanglement entropy is then given by
\begin{equation}
\label{Sent_an2}
\overline{\mathcal{S}}_{\textrm{ent}}(t)=\sum_{\ell=2}^\infty\sum_{d=1}^{\ell-1}
\frac{\mathcal{S}_{\textrm{ent}}^{(d,\ell)}(t)}{2^{\ell+1}}\leq
\sum_{\ell=2}^\infty\sum_{d=1}^{\ell-1}\frac{\ln \Gamma_d}{2^{\ell+1}}.
\end{equation}
Here $\mathcal{S}_{\textrm{ent}}^{(d,\ell)}(t)$ is the entanglement
entropy of a block of length $d$ obtained by tracing out $d-\ell$
sites in a segment of length $\ell$ and $\Gamma_d$ is the dimension of
the reduced density matrix of this block. For the \XX\ and the
hard-core boson case, in particular, we have $\Gamma_d=2^d$ leading to
the upper bound $\overline{\mathcal{S}}_{\textrm{ent}}(t)\leq \ln 2$.
A similar bound also exists for the Bose-Hubbard model in the general
case.

Within the purification approach used in the paper for the LCRG
calculations, the time-evolved state is a superposition of states for
each disorder configuration, i.e. in the non-interacting cases at
infinite disorder, different segment lengths are participating in the
superposition with the appropriate weight. Since states containing
long segments are exponentially suppressed, the entanglement entropy
with ancillas, $S_{\textrm{ent}}$, shown for the \XX\ case in the
inset of Fig.~2 in the paper, is still bounded. The existence of a
bound can be shown analytically in this case as well.

\section{II. Spin-only entanglement entropy for the \XXZ\ model}
To show that the two entanglement entropies defined in
Eq.~\eqref{Sent1} and Eq.~\eqref{Sent2}, respectively, have the same
functional dependence on $t$ for the \XXZ\ model we present in
Fig.~\ref{fig:exactee} the spin-only entanglement entropy defined in
Eq.~\eqref{Sent1}---which cannot easily be obtained using LCRG---for
finite systems of size $N=8,10,12$ where the disorder average has been
performed exactly. To make the dependence
$\overline{\mathcal{S}}_{\textrm{ent}}\sim \ln t$ at long times clearly visible,
the data are calculated with a time step of $\delta t=1$ and a
subsequent moving average with range $10$ is performed.

\begin{figure}[ht!]
  \centering
  \includegraphics[width=0.65\columnwidth]{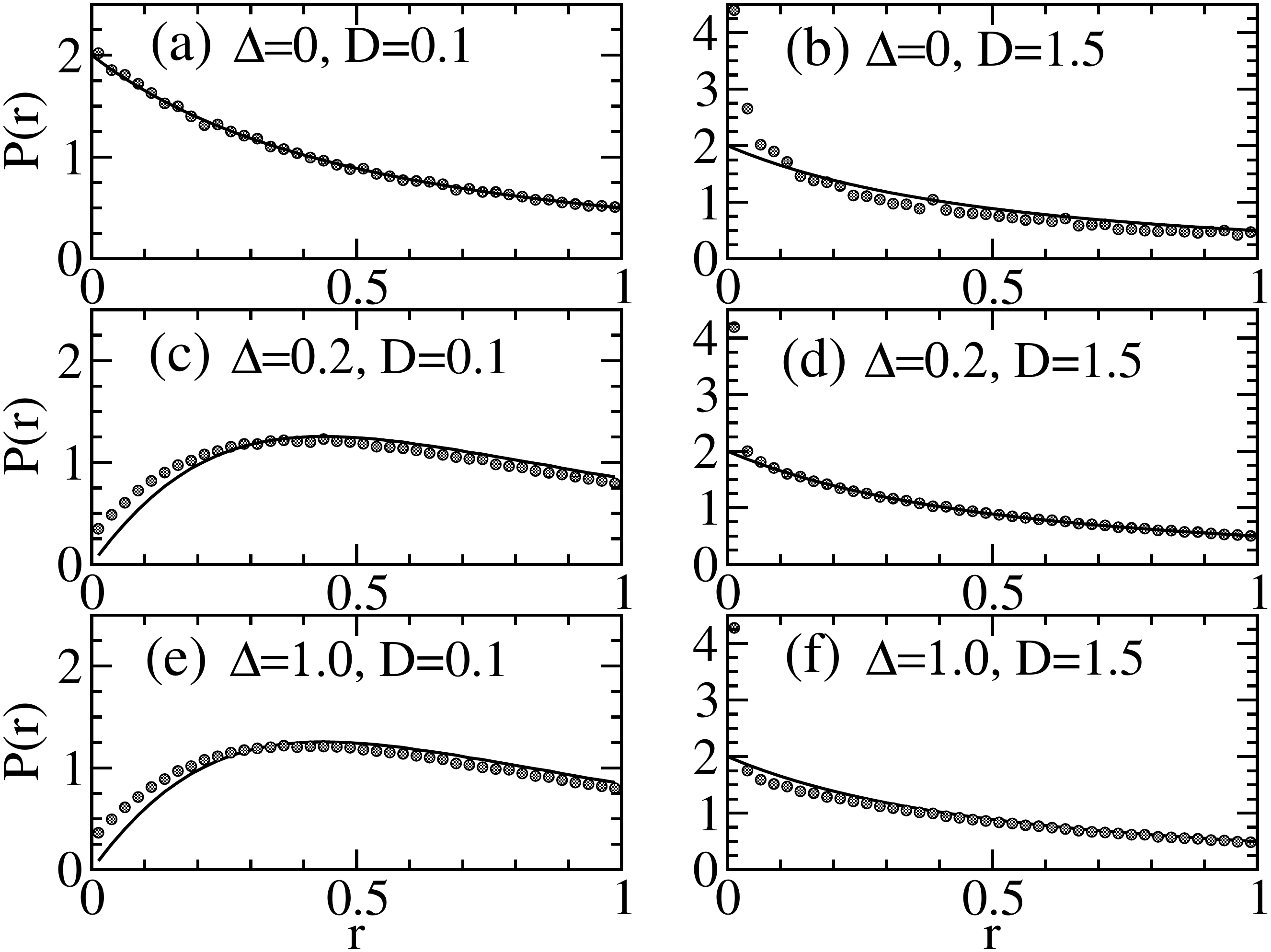}
  \caption{Gap distribution function $P(r)$ for \XXZ\ chains with
    $N=14$ sites (symbols). Lines represent $P(r)=2/(1+r)^2$ in the
    Poissonian case, or the result from diagonalizing random matrices
    in the GOE case. (a,b) In the non-interacting case, $\Delta=0$,
    the level statistics is Poissonian for all disorder strengths.
    (c,d) In the weakly interacting case, the level statistics changes
    from GOE at small disorder to Poissonian at intermediate disorder.
    (e,f) A change from GOE to Poissonian statistics is also observed
    in the strongly interacting case.}
  \label{fig:rss}
\end{figure}

\begin{figure}[ht!]
  \centering
  \includegraphics[width=0.65\columnwidth]{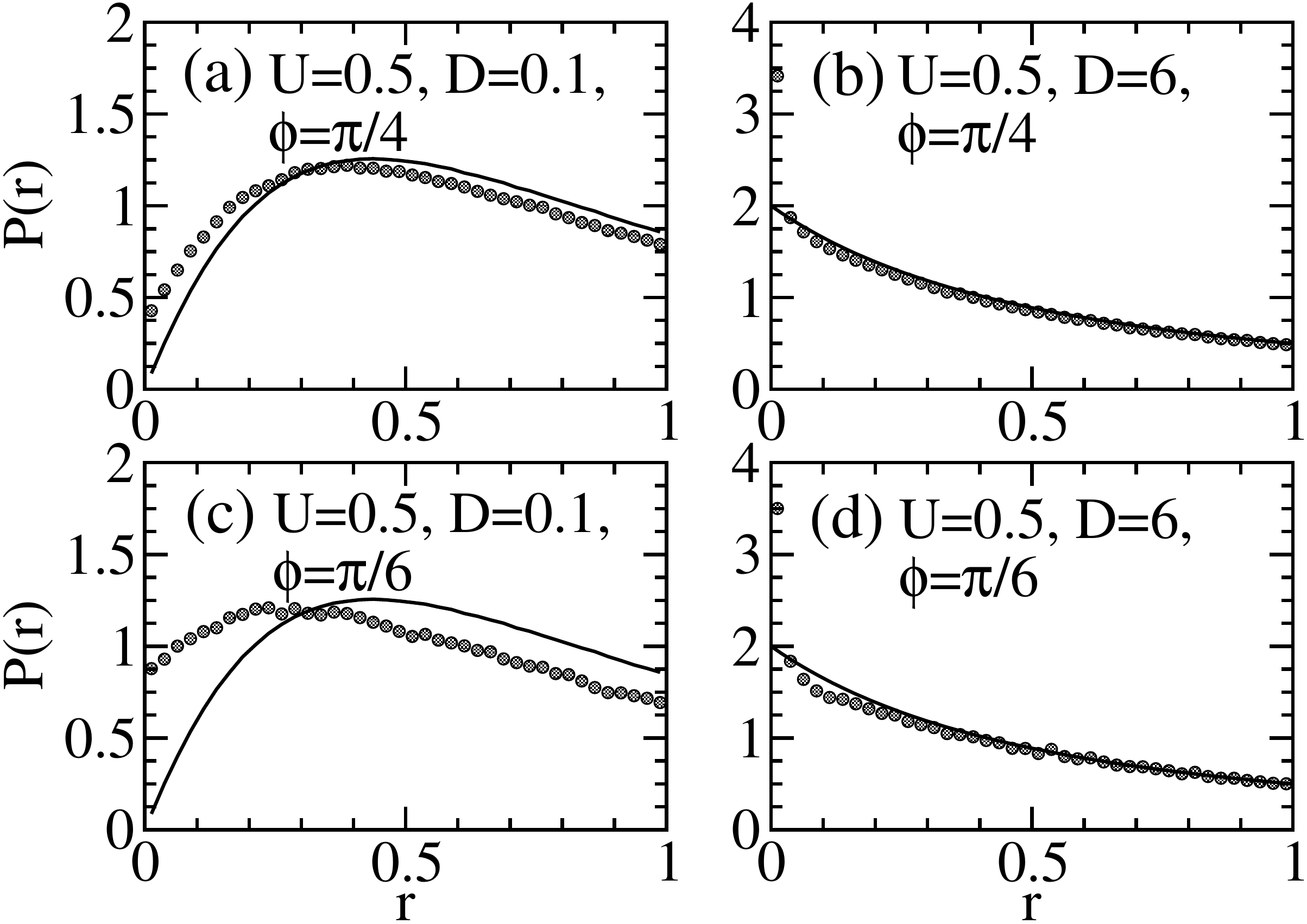}
  \caption{Gap distribution function $P(r)$ for Bose-Hubbard chains
    with $V=0$ and $N=12$ sites (symbols). Lines represent the
    theoretically expected distributions. The level statistics changes
    from GOE for weak disorder, panel (a,c), to Poissonian for
    intermediate disorder, panel (b,d). We consider two filling
    fractions for the immobile bosons parametrized by $\phi$, see
    text.}
  \label{fig:rss2}
\end{figure}

\section{III. Level statistics}
\label{Lev_stat}
The existence of an MBL phase for the Bose-Hubbard and the \XXZ\ model
with binary disorder is supported by the level statistics for finite
chains. In the paper we have been following the approach in
Refs.~\cite{oganesyan2007, pal2010} by considering the ratio $0\leq
r=\min\{\delta_n,\delta_{n-1}\}/\max\{\delta_n,\delta_{n-1}\}\leq 1$
between two consecutive gaps of adjacent energy levels,
$\delta_n=E_{n+1}-E_n\geq 0$. If a set of local or quasi-local
conserved charges exists which allows to uniquely label the
eigenvalues then the level statistics is uncorrelated and the
distribution function of gap ratios $r$ is given by $P(r)=2/(1+r)^2$.
We expect this to be the case if the model is integrable and when the
model is in an MBL phase. For a generic non-integrable model with
extended states, on the other hand, the level statistics is expected
to follow a Gaussian orthogonal ensemble (GOE). In this case, the
distribution function of gap ratios $P(r)$ is not known analytically.
As in Ref.~\cite{oganesyan2007} we determine $P(r)$ by exactly
diagonalizing $10000$ random $(3432\times 3432)$ GOE
matrices.

\subsection{\XXZ\ model}

The \XXZ\ chain is integrable in the clean case leading to a
Poissonian level statistics. As shown in Fig.~\ref{fig:rss}(a,b) the
level statistics remains Poissonian also with disorder in the \XX\
case, $\Delta=0$, where the model is equivalent to free spinless
fermions.

For finite interaction strength, on the other hand, we find a GOE
statistics for weak disorder and a Poissonian statistics for
intermediate disorder, see Fig.~\ref{fig:rss}(c,d), and
Fig.~\ref{fig:rss}(e,f). In the intermediate disorder case, the
Poissonian statistics is already clearly visible which means that the
localization length is smaller than the system size, strongly
supporting the existence of an MBL phase. For the weakly disordered
case, shown in Fig.~\ref{fig:rss}(c) and Fig.~\ref{fig:rss}(e), there
are two possibilities: the system either has extended states or the
localization length is much larger than the considered system size.
Finding out which one of the two scenarios is true is not easily
possible based on finite size data and is beyond the scope of this
letter. The crossover from a Poissonian level statistics in the clean
integrable case to a GOE statistics at weak disorder can also be
understood as the onset of chaos which might remain incomplete because
of the formation of new quasi-local charges \cite{YurovskyOlshanii},
see also Fig.~4 in the main paper. The deviations in
Fig.~\ref{fig:rss}(b,d,f) near $r\approx 0$ from the result expected
for a Poissonian statistics are finite size effects and become smaller
with increasing system size.

\subsection{Bose-Hubbard model with $V=0$}
Finally, we present additional supporting data for the level
statistics of the Bose-Hubbard model with $V=0$ in
Fig.~\ref{fig:rss2}.
The Bose-Hubbard model is not integrable so that the level statistics
is GOE without disorder. For small disorder we still find agreement with GOE
statistics, see Fig.~\ref{fig:rss2}, which means, in analogy to the
\XXZ\ case, that the states either remain extended or that the
localization length for this disorder strength is much larger than the
system size. For disorder strength $D=6$ shown in
Fig.~\ref{fig:rss2}(b), on the other hand, the level statistics is
Poissonian and the system in an MBL phase. 

\begin{figure}[ht!]
  \centering
  \includegraphics[width=0.6\columnwidth]{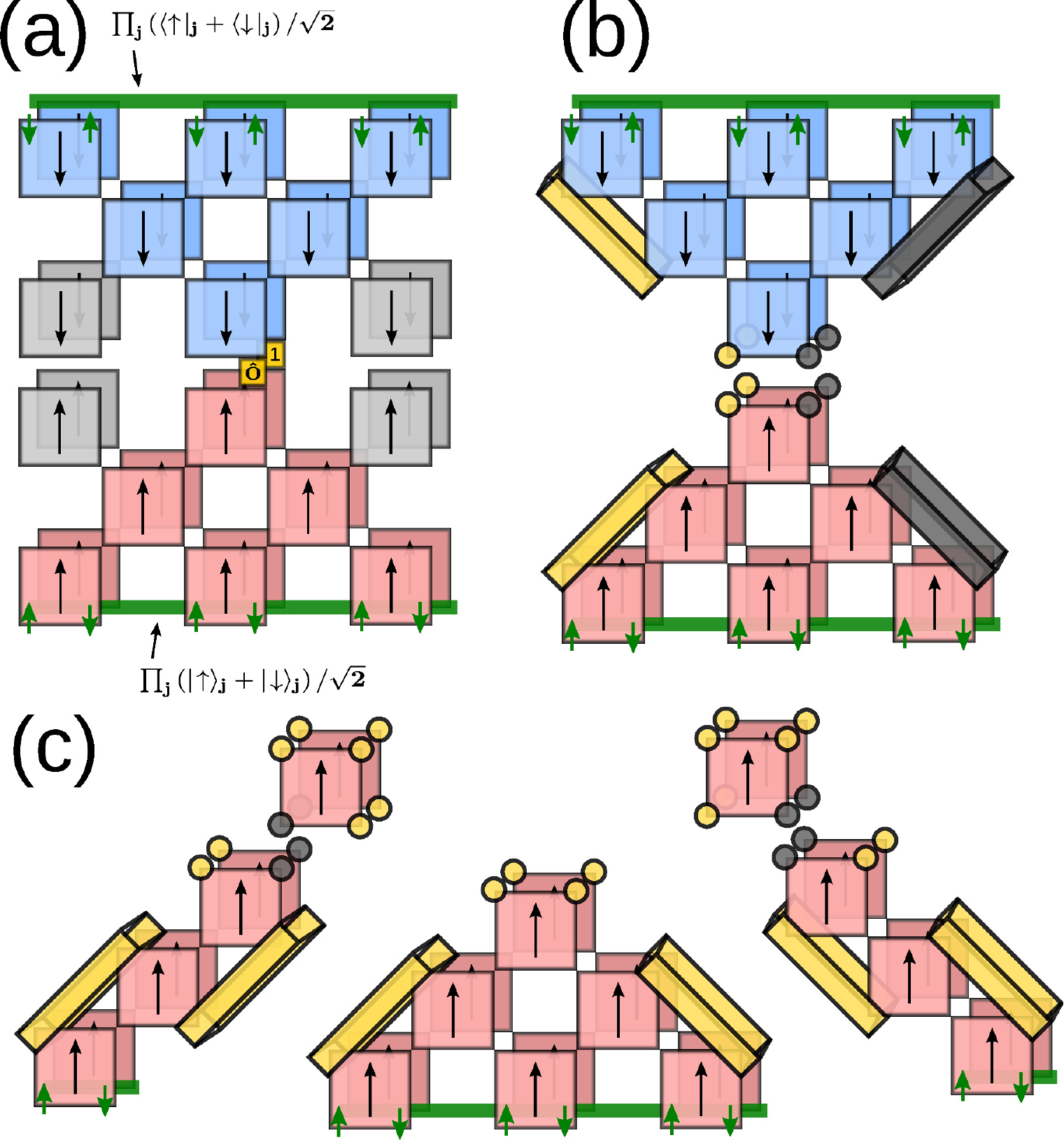}
  \caption{LCRG algorithm with exact disorder average for the \XXZ\
    model: (a) Measurement of a one-point function
    $\langle\hat{O}(t)\rangle$. The grey plaquettes cancel out due to
    unitarity. (b) Reduced density matrix for renormalization. Grey
    indices are summed over. Circles denote single spin {\it or}
    ancilla indices, boxes the block indices of spins {\it and}
    ancillas. (c) Growth step of left and right transfer matrices, and
    central light cone. }
  \label{fig9}
\end{figure}

\section{IV. Details on the LCRG algorithm}

We study the dynamics of the disordered \XXZ\ and Bose-Hubbard models
with the LCRG algorithm \cite{enss2012lcrg}. This variant of
time-dependent DMRG makes use of the light cone structure inherent to
the Trotter-Suzuki decomposition of the time evolution operator, as
shown in Fig.~\ref{fig9}. If the Trotter step is chosen small enough
with respect to the Lieb-Robinson velocity of the system, simulating
the depicted light cone is sufficient to calculate observables in the
thermodynamic limit accurately. To simulate binary disorder, we follow
the purification approach introduced in Ref.~\onlinecite{paredes2005}:
to each lattice site we add an ancilla spin-$1/2$. The ancilla degrees
of freedom are coupled to the physical system by adding to the
Hamiltonian $H^{dis}=-2JD\sum_j s^z_j s^z_{j,anc}$ for the \XXZ\
model, and $H^{dis}=JD\sum_j n_j^A n_j^B$ for the Bose-Hubbard model,
where $n_j^A$ ($n_j^B$) is the particle number operator for the mobile
(immobile) bosons. For binary disorder, we prepare the ancilla spins
(immobile bosons) in the initial product state $\prod_j
(|\uparrow\rangle_j + |\downarrow\rangle_j)/\sqrt{2}$ or
$\prod_j(|1\rangle_j+|0\rangle_j)/\sqrt{2}$, respectively, so they
represent a uniform superposition of all possible realizations of
static binary disorder.  We note that it is possible to prepare the
immobile bosons in the more general state $\prod_j
(\sin\phi|1\rangle_j+\cos\phi|0\rangle_j)$, for any $\phi\in[0,2\pi)$,
instead, realizing a weighted superposition of binary disorder. Here,
$\sin^2\phi$ determines the average density $\bar{n}^B$ of the
immobile boson species in an experiment. While a thorough analysis of
the stability of the MBL phase with respect to $\bar{n}^B$ is outside
the scope of this work, we present a comparison of the level
statistics the Bose-Hubbard model for average densities
$\bar{n}^B=1/2$ $(\phi=\pi/4)$ and $\bar{n}^B=1/4$ $(\phi=\pi/6)$ in
Fig.~\ref{fig:rss2}.  The data demonstrate that the MBL phase is
stable against a variation in the density of the immobile bosons.

\begin{figure}[ht!]
  \centering
  \includegraphics[width=0.5\columnwidth]{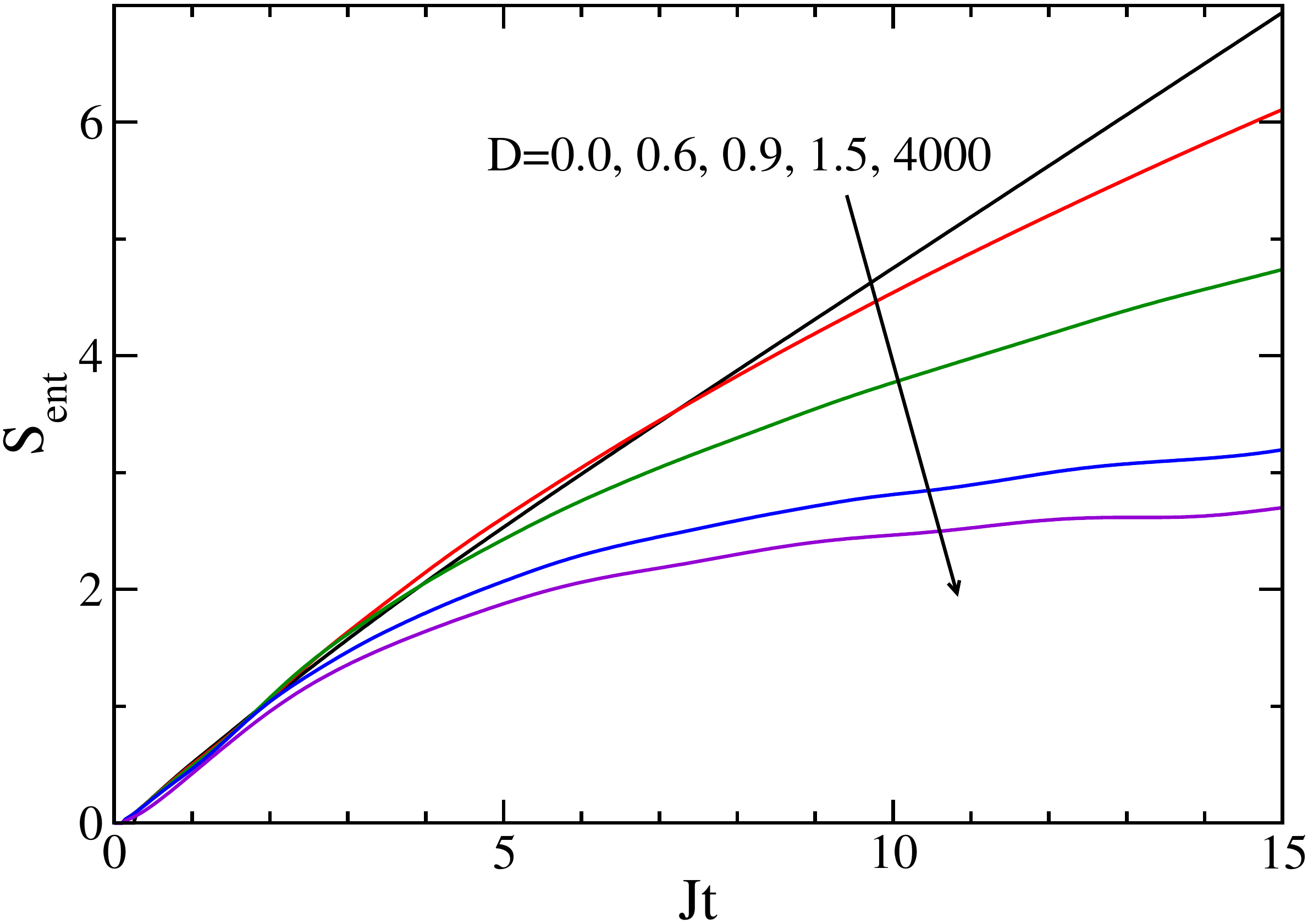}
  \caption{Entanglement entropy for the \XXZ\ model with $\Delta=1$.}
  \label{fig10}
\end{figure}

Details of the LCRG algorithm with ancillas are shown in
Fig.~\ref{fig9}: in each step, the left and right transfer matrices
are grown by adding a single Trotter plaquette (Fig.~\ref{fig9}c).
Then, the central light cone is alternatingly grown to the left and
right by contracting it with the corresponding transfer matrix.
Subsequently, a standard DMRG renormalization is performed on the
reduced density matrix depicted in Fig.~\ref{fig9}b. A defining
feature for the MBL phase is the logarithmic growth of the
entanglement entropy, $S_{ent}\propto\ln t$. This makes it possible to
simulate the disordered system for times up to an order of magnitude
longer than the clean system, in spite of having doubled the local
Hilbert space dimension. In Fig.~\ref{fig10}, we show the gradual
improvement of the scaling of the entanglement entropy with increasing
disorder strength on the example of the Heisenberg chain.


\begin{thebibliography}{49}
\bibitem{BlochDalibard}
I.~Bloch, J.~Dalibard, and W.~Zwerger, Rev. Mod. Phys. \textbf{80}, 885 (2008).
\bibitem{GreinerMandel}
M.~Greiner \textit{et~al.}, Nature (London) 
  \textbf{415}, 39 (2002).
\bibitem{par04}
B.~Paredes \textit{et~al.}, Nature (London) \textbf{429}, 277 (2004).
\bibitem{KinoshitaWenger}
T.~Kinoshita, T.~Wenger, and D.~S. Weiss, Nature (London) \textbf{440}, 900 (2006).
\bibitem{anderson1958}
P.~W. Anderson, Phys.\ Rev. \textbf{109}, 1492 (1958).
\bibitem{RoatiDerricoBillyJosseJendrzejewskiBernardWhitePasienski}
G.~Roati \textit{et~al.}, Nature (London) \textbf{453}, 895 (2008); 
J.~Billy \textit{et~al.}, Nature (London) \textbf{453}, 891
  (2008);
M.~White \textit{et~al.},
  Phys. Rev. Lett. \textbf{102}, 055301 (2009);
F.~Jendrzejewski \textit{et~al.}, Nat. Phys.
  \textbf{8}, 398 (2012).
\bibitem{GavishCastinMorrisonKantianRoscildeCirac}
U.~Gavish and Y.~Castin, Phys. Rev. Lett. \textbf{95}, 020401 (2005);
T.~Roscilde and J.~I. Cirac, Phys. Rev. Lett. \textbf{98}, 190402 (2007);
S.~Morrison \textit{et~al.}, New J. Phys. \textbf{10}, 073032 (2008).
\bibitem{HorstmannCiracHorstmannDuerr}
B.~Horstmann, J.~I. Cirac, and T.~Roscilde, Phys. Rev. A \textbf{76}, 043625
  (2007);
B.~Horstmann, S.~D{\"u}rr, and T.~Roscilde, Phys. Rev. Lett. \textbf{105},
  160402 (2010).
\bibitem{OspelkausOspelkaus_localization}
S.~Ospelkaus \textit{et~al.}, Phys. Rev. Lett. \textbf{96}, 180403 (2006).
\bibitem{basko2006}
D.~M. Basko, I.~L. Aleiner, and B.~L. Altshuler, Ann.\ Phys.\ (NY)
  \textbf{321}, 1126 (2006).
\bibitem{oganesyan2007}
V.~Oganesyan and D.~A. Huse, Phys.\ Rev.~B \textbf{75}, 155111 (2007).
\bibitem{pal2010}
A.~Pal and D.~A. Huse, Phys.\ Rev.~B \textbf{82}, 174411 (2010).
\bibitem{monthus2010canovi2011}
C.~Monthus and T.~Garel, Phys.\ Rev.~B \textbf{81}, 134202 (2010);
E.~Canovi \textit{et~al.}, Phys.\ Rev.~B
  \textbf{83}, 094431 (2011);
J.~Z. Imbrie, arXiv:1403.7837;
T.~Grover, arXiv:1405.1471.
\bibitem{1D_case} 
We concentrate here on the one-dimensional case but the argument is general and also applicable in higher dimensions.
\bibitem{RigolDunjkoPRL}
M.~Rigol, V.~Dunjko, V.~Yurovsky, and M.~Olshanii, Phys. Rev. Lett.
  \textbf{98}, 050405 (2007).
\bibitem{vosk2013}
R.~Vosk and E.~Altman, Phys.\ Rev.\ Lett. \textbf{110}, 067204 (2013).
\bibitem{serbyn2013localhuse2013} C.~Gogolin, M.~P.~M\"uller, and
  J.~Eisert, Phys.\ Rev.\ Lett. \textbf{106}, 040401 (2011);
  G.~Carleo, F.~Becca, M.~Schir\'o, and M.~Fabrizio, Sci.~Rep.
  \textbf{2}, 243 (2012); M.~Serbyn, Z.~Papi{\'c}, and D.~A. Abanin,
  Phys.\ Rev.\ Lett. \textbf{111}, 127201 (2013); D.~A. Huse and
  V.~Oganesyan, arXiv:1305.4915; Y.~Bar~Lev and D.~R. Reichman, Phys.\
  Rev.\ B \textbf{89}, 220201(R) (2014); R.~Vasseur,
  S.~A.~Parameswaran, and J.~E.~Moore, arXiv: 1407.4476.
\bibitem{ProsenSirkerKonstantinidis}
T.~Prosen, Phys. Rev. Lett. \textbf{106}, 217206 (2011);
J.~Sirker, N.~P. Konstantinidis, F.~Andraschko, and N.~Sedlmayr, Phys. Rev. A
  \textbf{89}, 042104 (2014).
\bibitem{dechiara2006znidaric2008vosk2013dynamicalserbyn2014}
G.~De~Chiara, S.~Montangero, P.~Calabrese, and R.~Fazio, J.~Stat.\ Mech.
  P03001 (2006);
M.~{\v{Z}}nidari{\v{c}}, T.~Prosen, and P.~Prelov{\v{s}}ek, Phys.\ Rev.~B
  \textbf{77}, 064426 (2008);
R.~Vosk and E.~Altman, Phys.\ Rev.\ Lett.\ \textbf{112}, 217204 (2014);
A.~Nanduri \textit{et~al.}, Phys.\ Rev.\ B \textbf{90}, 064201;
M.~Serbyn \textit{et~al.}, arXiv:1403.0693.
\bibitem{bardarson2012}
J.~H. Bardarson, F.~Pollmann, and J.~E. Moore, Phys.\ Rev.\ Lett. \textbf{109},
  017202 (2012).
\bibitem{bravyi2006}
S.~Bravyi, M.~B. Hastings, and F.~Verstraete, Phys.\ Rev.\ Lett. \textbf{97},
  050401 (2006).
\bibitem{igloi2012}
F.~Igl{\'o}i, Z.~Szatm{\'a}ri, and Y.-C. Lin, Phys.\ Rev.~B \textbf{85}, 094417
  (2012).
\bibitem{AltmanVoskReview}
E.~Altman and R.~Vosk, arXiv: 1408.2834 (2014).
\bibitem{TrotzkyChen}
S.~Trotzky \textit{et~al.}, Nat. Phys. \textbf{8}, 325 (2012).
\bibitem{paredes2005}
B.~Paredes, F.~Verstraete, and J.~I. Cirac, Phys.\ Rev.\ Lett. \textbf{95},
  140501 (2005).
\bibitem{SupplMat}
See Supplemental Material for exact results in free models, entanglement entropy, level
statistics for XXZ and Bose-Hubbard models, and a description of the
LCRG algorithm.
\bibitem{barmettler2009barmettler2010}
P.~Barmettler \textit{et~al.}, Phys.\ Rev.\
  Lett. \textbf{102}, 130603 (2009);
  New J. Phys.
  \textbf{12}, 055017 (2010).
\bibitem{enss2012lcrg}
T.~Enss and J.~Sirker, New J. Phys. \textbf{14}, 023008 (2012).
\bibitem{GiulianoRossini}
D.~Giuliano, D.~Rossini, P.~Sodano, and A.~Trombettoni, Phys. Rev. B
  \textbf{87}, 035104 (2013).
\bibitem{FukuharaKantian}
T.~Fukuhara \textit{et~al.}, Nat. Phys.
  \textbf{9}, 235 (2013).
\bibitem{Hazzard}
N.~Y. Yao \textit{et~al.}, arXiv:1311.7151;
K.~R.~A. Hazzard \textit{et~al.}, arXiv:1406.0937.
\bibitem{white1992} S.~R. White, Phys.\ Rev.\ Lett. \textbf{69}, 2863
  (1992); A.~J.~Daley \textit{et al.}, J.~Stat.~Mech.~P04005 (2004);
  S.~R.~White and A.~E.~Feiguin, Phys.\ Rev.\ Lett.\ \textbf{93},
  076401 (2004).
\bibitem{lieb1972}
E.~H. Lieb and D.~W. Robinson, Commun.\ Math.\ Phys. \textbf{28}, 251 (1972).
\bibitem{AndraschkoSirker}
F.~Andraschko and J.~Sirker, Phys. Rev. B \textbf{89}, 125120 (2014).
\bibitem{YurovskyOlshanii}
V.~A.~Yurovsky, and M. Olshanii, Phys. Rev. Lett. \textbf{106}, 025303 (2011).
\bibitem{MotoyamaEisaki}
N.~Motoyama, H.~Eisaki, and S.~Uchida, Phys. Rev. Lett. \textbf{76}, 3212
  (1996).
\bibitem{Kojima}
K.~M. Kojima \textit{et~al.}, Phys. Rev. B
  \textbf{70}, 094402 (2004).
\bibitem{SirkerLaflorencieSirkerLaflorencie2}
J.~Sirker \textit{et~al.}, Phys. Rev.
  Lett. \textbf{98}, 137205 (2007);
J.~Stat.\ Mech. P02015 (2008).
\bibitem{EnssAndraschkoSirker_prep}
T.~Enss, F.~Andraschko, and J.~Sirker (2014) (to be published).
\end{thebibliography}
\end{document}